\documentclass[10pt, twocolumn, twoside]{IEEEtran}

\usepackage{cite}
\usepackage{enumerate}
\usepackage{graphicx}
\usepackage[cmex10]{amsmath} % prevents amsmath from using a Type 3 font for math within footnotes
\usepackage{amssymb}
\usepackage{mathtools}
\usepackage{xspace}
\usepackage{subfigure}
\usepackage[lined,linesnumbered,ruled,commentsnumbered]{algorithm2e} % for algorithm
\usepackage{colonequals}
\usepackage[mathscr]{euscript}
\usepackage{fixltx2e}    % for text subscript
\usepackage{amsthm}
\usepackage{mathrsfs}

\usepackage{epsfig}
\usepackage{epstopdf}

\usepackage{tikz}
\usepackage{pgfplots}

\usetikzlibrary{arrows,shapes,backgrounds,plotmarks,positioning}

\pgfplotsset{compat=newest}                         % move axis labels close to the tick label automatically
\pgfplotsset{plot coordinates/math parser=false}
\newlength\figureheight
\newlength\figurewidth

\newtheorem{theorem}{Theorem}[section]
\newtheorem{lemma}[theorem]{Lemma}
\newtheorem{definition}[theorem]{Definition}
\newtheorem{corollary}[theorem]{Corollary}

\newtheorem{example}[theorem]{Example}

\newcommand{\op}{\text}
\newcommand{\RZ}[1]{\mathsf{Z}_{#1}}
\newcommand{\RW}[1]{\mathsf{W}_{#1}}
\newcommand{\RRCO}{\mathscr{R}_{\op{CO}}}
\newcommand{\RCO}{R_{\op{CO}}}
\newcommand{\Card}[1]{|#1|}
\newcommand{\Set}[1]{\{#1\}}
\newcommand{\Core}{\mathscr{C}}
\newcommand{\Has}[1]{W_{#1}}
\newcommand{\Pat}{\mathcal{P}}

\newcommand{\Vu}{f_{\alpha}}
\newcommand{\V}[1]{f_{#1}}
\newcommand{\VuHat}{\hat{f}_{\alpha}}       %
\newcommand{\VHat}[1]{\hat{f}_{#1}}
\newcommand{\VuHash}{f^\#_{\alpha}}
\newcommand{\VHash}[1]{f^\#_{#1}}
\newcommand{\VuHashHat}{\hat{f}^\#_{\alpha}}
\newcommand{\VHashHat}[1]{\hat{f}^\#_{#1}}

\newcommand{\Real}{\mathbb{R}}
\newcommand{\RealP}{\mathbb{R}_{+}}    % nonnegative real number
\newcommand{\N}{\mathbb{N}_0}          % nonnegative integer number
\newcommand{\Z}{\mathbb{Z}}            % integer number
\newcommand{\F}{\mathbb{F}}            % field

\newcommand{\fgref}[1]{Fig.~\ref{#1}}       % figure reference

 {\begin{list}{}%
         {\setlength{\leftmargin}{#1}}%
         \item[]%
 }
 {\end{list}}

\begin{document}

\title{A Game-theoretic Perspective on Communication for Omniscience}

\author{Ni~Ding\IEEEauthorrefmark{1}, Chung~Chan\IEEEauthorrefmark{2}, Tie~Liu\IEEEauthorrefmark{3}, Rodney~A.~Kennedy\IEEEauthorrefmark{1}, Parastoo~Sadeghi\IEEEauthorrefmark{1}

\thanks{\IEEEauthorblockA{\IEEEauthorrefmark{1}Ni Ding, Rodney A. Kennedy and Parastoo Sadeghi are with the Research School of Engineering, College of Engineering and Computer Science, the Australian National University (email: $\{$ni.ding, rodney.kennedy, parastoo.sadeghi$\}$@anu.edu.au).   }}
\thanks{\IEEEauthorblockA{\IEEEauthorrefmark{2}Chung Chan is with the Institute of Network Coding at Chinese University of Hong Kong (email: cchan@inc.cuhk.edu.hk). }}
\thanks{\IEEEauthorblockA{\IEEEauthorrefmark{3}Tie Liu is with the Department of Electrical and Computer Engineering, Texas A\&M University (email: tieliu@tamu.edu). }}
}

%\markboth{journal name}{Ding \MakeLowercase{\textit{et al.}}: Iterative Merging Algorithm for Successive Packet Recovery in Coded Cooperative Data Exchange}

\maketitle

\begin{abstract}
We propose a coalition game model for the problem of communication for omniscience (CO). In this game model, the core contains all achievable rate vectors for CO with sum-rate being equal to a given value. Any rate vector in the core distributes the sum-rate among users in a way that makes all users willing to cooperate in CO. We give the necessary and sufficient condition for the core to be nonempty. Based on this condition, we derive the expression of the minimum sum-rate for CO and show that this expression is consistent with the results in multivariate mutual information (MMI) and coded cooperative data exchange (CCDE). We prove that the coalition game model is convex if the sum-rate is no less than the minimal value. In this case, the core is non-empty and a rate vector in the core that allocates the sum-rate among the users in a fair manner can be found by calculating the Shapley value.
\end{abstract}

%\begin{IEEEkeywords}
%coalitional game, communication for omniscience (CO), multivariate mutual information (MMI).
%\end{IEEEkeywords}

\section{introduction}

Communication for omniscience (CO) is a problem proposed in \cite{Csiszar2004}. It is assumed that there is a group of users in the system and each of them observes a component of a discrete memoryless multiple source in private. The users can exchange their information in certain way, e.g., communicating over lossless broadcast channels, so as to attain \textit{omniscience}, the state that each user obtains the total information in the entire multiple source in the system. The CO problem in \cite{Csiszar2004} is based on an asymptotic source model. The coded cooperative data exchange (CCDE) problem proposed in \cite{Roua2010} is a special case of the CO problem where the source model is a non-asymptotic finite linear packet one. The non-asymptotic model differs from the asymptotic one in that the communication rates only take integer values. By allowing packet splitting, the CCDE problem has been extended for asymptotic model in \cite{CourtIT2014,Ding2015ISIT}. Independently, the same model has been considered in the closely related secret key agreement (SKA) problem by \cite{Chan2011ITW}, and is called the finite linear source model.

Determining the minimum sum-rate and finding an optimal rate vector that allocates the minimum sum-rate such that omniscience is achievable are two fundamental problems in CO. The expressions of the minimum sum-rate for asymptotic and non-asymptotic models are derived in \cite{ChanSuccessiveIT} based on multivariate mutual information (MMI) for SKA \cite{ChanMMI} and in \cite{Ding2015ISIT} for CCDE. It is also shown that an optimal rate vector can be solved via submodular function minimization (SFM) algorithms in strongly polynomial time, e.g., the algorithms in \cite{CourtIT2014,MiloIT2015} based on Edmond's greedy algorithm \cite{Edmonds2003Convex}. However, these works only focus on finding a solution for CO while neglecting the users' motivation to cooperate. For example, the algorithms in \cite{CourtIT2014,MiloIT2015} usually output an unfair rate vector which may discourage some users from taking part in CO.

In this paper, we view the CO problem from a coalitional game theoretic perspective. In this game model, each user is assumed to be self-determined in that they can decide whether or not to cooperate and join a certain coalition, a subset of the users. The core of the game is the set of achievable rate vectors with sum-rate being equal to a given value in CO and any rate vector in the core distributes the sum-rate among users in a way that makes all users willing to cooperate in the grand coalition, the entire user set. By using the concepts and related results of submodularity and its base polyhedron \cite{Fujishige2005}, we derive a necessary and sufficient condition for the core to be nonempty. We give the expressions of the minimum sum-rate for asymptotic and non-asymptotic models and show that they are in agreement with the results in \cite{ChanSuccessiveIT,Ding2015ISIT}. The coalitional game model also addresses another problem in CO: how to allocate the rate fairly to motivate the users to cooperate in asymptotic model. We show that the game is equivalent to a convex game and a fair rate allocation method can be determined by Shapley value if the sum-rate is no less than the minimum value. As compared to the existing method that addresses the fairness in CCDE in \cite{Taj2011}, the main advantage of Shapley value is that it can be calculated in a decentralized manner, i.e., it is possible for each user to obtain his/her tuple in Shapley value by him/herself.

\section{System Model}
\label{sec:system}

Let $V$ be a finte set. We assume that there are $\Card{V}>1$ users in the system. A random variable will be denoted by the san serif font as in $\mathsf{Z}$, and its alphabet by the usual math font as in $Z$. $\RZ{V}=(\RZ{i}:i\in V)$ is a vector of discrete random variables indexed by $V$. For each $i\in V$, user $i$ can privately observe an $n$-sequence $\RZ{i}^n$ of the random source $\RZ{i}$ that is i.i.d.\ generated according to the joint distribution $P_{\RZ{V}}$. We allow users exchange their sources directly so as to let all users $i\in V$ recover the source sequence $\RZ{V}^n$. We consider both asymptotic and non-asymptotic models. In the asymptotic model, we will characterize the asymptotic behavior as the \emph{block length} $n$ goes to infinity. In non-asymptotic model, the communication rates are required to be integer valued.

Let $r_V=(r_i:i\in V)$ be a rate vector indexed by $V$. We call $r_V$ an achievable rate vector if omniscience is possible by letting users communicate with the rates designated by $r_V$. Let $r(V)=\sum_{i\in V} r_i$. For $X,Y \subseteq V$, let $H(\RZ{X})$ be the amount of randomness in $\RZ{X}$ measured by Shannon entropy \cite{YeungITBook}
% $$ H(\RZ{X}) \equiv E \Big[ \log_2\frac{1}{P_{\RZ{V}}} \Big] = \sum_{z_X\in Z_X}\log_2\frac{1}{P_{\RZ{X}}(z_X)}$$
and $H(\RZ{X}|\RZ{Y})=H(\RZ{X \cup Y})-H(\RZ{Y})$ be the conditional entropy of $\RZ{X}$ given $\RZ{Y}$. It is shown in \cite{Csiszar2004} that an achievable rate must satisfy the Slepian-Wolf constraints:
\begin{equation} \label{eq:SWConstrs}
    r(X) \geq H(\RZ{X}|\RZ{V\setminus X}), \quad \forall X \subset V.
\end{equation}
The interpretation of the Slepian-Wolf constraint on $X$ is: To achieve CO, the total amount of information sent from user set $X$ should be at least complementary to total amount of information that is missing in user set $V \setminus X$. The set of all achievable rate vectors is
$$ \RRCO(\RZ{V})=\Set{ r_V\in\RealP^{|V|} \colon r(X) \geq H(\RZ{X}|\RZ{V\setminus X}),\forall X \subset V } $$
and the minimum sum-rate can be determined by the following linear programming (LP)
$$ \RCO(\RZ{V})=\min\Set{r(V) \colon r_V\in \RRCO(\RZ{V})}. $$
We denote the set of optimal rates as
\begin{equation}	
  \RRCO^*(\RZ{V})=\Set{r_V\in \RRCO(\RZ{V}) \colon r(V)=\RCO(\RZ{V})}.\label{eq:RRCO*}
\end{equation}
For non-asymptotic model, the achievable rate set is $\RRCO(\RZ{V}) \cap \Z^{|V|}$, $\RCO(\RZ{V})\in\N$ and the optimal rate set is $\RRCO^*(\RZ{V})\cap\Z^{|V|}$.

In CCDE, it is assumed that user $i$ obtains a packet set $\Has{\{i\}}$ that contains finite number of packets each of which belongs to a field $\F_q$. The users transmit linear combinations of their packet set via lossless wireless channels so as to help the others to recover all packets in $\Has{V}=\cup_{i\in V} \Has{\{i\}}$. In this problem, the value of entropy function $H(\RZ{X})$ can be obtained by counting the number of packets in $\Has{X}$, i.e., $H(\RZ{X})=|\Has{X}|$ and $H(\RZ{X}|\RZ{Y})=|\Has{X \cup Y}|-|\Has{Y}|$, and all results derived in this paper hold.

\begin{example} \label{ex:main}
Consider the set $V=\{1,2,3\}$ where $3$ users observe respectively
    \begin{align}
        \RZ{1} &= (\RW{a},\RW{b},\RW{c},\RW{d},\RW{e}),   \nonumber\\
        \RZ{2} &= (\RW{a},\RW{b},\RW{f}),   \nonumber\\
        \RZ{3} &= (\RW{c},\RW{d},\RW{f}).   \nonumber
    \end{align}
$\RW{i}$'s are independent uniformly distributed random bits. The users exchange their private observations to achieve the global omniscience of $\RZ{V}=(\RW{a},\dotsc,\RW{f})$. In this system, the achievable rate set is
    \begin{align}
        \RRCO(\RZ{V})=\Set{ &r_V\in\RealP^{|V|} \colon r(\emptyset)=0,                           \nonumber \\
         &r(\{1\}) \geq H(\RZ{1}|\RZ{\{2,3\}})=1 ,   \nonumber\\
         &r(\{2\}) \geq H(\RZ{2}|\RZ{\{1,3\}})=0 ,   \nonumber\\
         &r(\{3\}) \geq H(\RZ{3}|\RZ{\{1,2\}})=0 ,   \nonumber\\
         &r(\{1,2\}) \geq H(\RZ{\{1,2\}}|\RZ{3})=3 ,   \nonumber\\
         &r(\{1,3\}) \geq H(\RZ{\{1,3\}}|\RZ{2})=3 ,   \nonumber\\
         &r(\{2,3\}) \geq H(\RZ{\{2,3\}}|\RZ{1})=1 }.   \nonumber
    \end{align}
It can be shown that $\RCO(\RZ{V})=3.5$ and $\RRCO^*(\RZ{V})=\{(2.5,0.5,0.5)\}$ for the asymptotic model and $\RCO(\RZ{V})=4$ and $\RRCO^*(\RZ{V})\cap\Z^{|V|}=\{(3,0,1),(2,1,1),(3,1,0)\}$ for the non-asymptotic model.
\end{example}

\section{Coalitional Game}

We model the system as a coalitional game $G(V,\alpha,\Vu)$. In this game, it is assumed that the users can choose to cooperate and form coalitions. A coalition is a group/set of clients that is denoted by $X \subseteq V$ and $V$ is called the \textit{grand coalition}. Let $\alpha\in\RealP$. We define the \textit{characteristic function} for a given value of $\alpha$ as
$$ \Vu(X)=\begin{cases} H(\RZ{X}|\RZ{V\setminus X}) & X \subset V \\ \alpha & X=V \end{cases}. $$
We call $\Vu(X)$ the \textit{value} of coalition $X$ which quantifies the payoff of forming coalition $X$.

\subsection{Core}

The \textit{core} of $G(V,\alpha,\Vu)$ is
\begin{multline} \label{eq:Core}
    \Core_G = \{ r_V\in\RealP^{|V|} \colon r(X) \geq \Vu(X),\forall X \subseteq V, \\ r(V)=\Vu(V) \}
\end{multline}
for asymptotic model and $\Core_G\cap\Z^{|V|}$ for non-asymptotic model. Let the polyhedron of the characteristic function $\Vu$ be
$$ P(\Vu,\geq)= \Set{r_V\in\RealP^{|V|} \colon r(X) \geq H(\RZ{X}|\RZ{V\setminus X}),\forall X \subseteq V}. $$
The core $\Core_G$ is exactly the base polyhedron of $\Vu$:
$$ B(\Vu,\geq)= \Set{r_V\in\RealP^{|V|} \colon r_V \in P(\Vu,\geq) \colon r(V)=\Vu(V) }. $$
Consider the constraints in $B(\Vu,\geq)$. Let
$$ \VuHash(X)=\Vu(V)-\Vu(V \setminus X)=\alpha-\Vu(V \setminus X), \forall X \subseteq V$$
%\begin{align}
%    \VuHash(X)&=\Vu(V)-\Vu(V \setminus X)  \nonumber \\
%                  &=\begin{cases} 0 & X=\emptyset \\ \alpha-H(\RZ{V\setminus X}|\RZ{X}) & \text{otherwise}  \end{cases}   \nonumber
%\end{align}
be the \textit{dual set function} of $f_\alpha$. If we restrict the rate vector $r_V$ to satisfy $r(X) \geq \Vu(X)$ for some $X\subseteq V$ and the sum-rate $r(V)=\alpha$, then we necessarily put constraint
$$r(V \setminus X)=r(V)-r(X) \leq \VuHash(V \setminus X) $$
on set $V \setminus X$. By converting the constraints in $B(f_\alpha,\geq)$ in the same way for all $X \subseteq V$, we get the base polyhedron
$$ B(f_\alpha^\#,\leq)= \Set{r_V\in\RealP^{|V|} \colon r_V \in P(f_\alpha^\#,\leq),r(V)=f_\alpha^\#(V) } $$
such that $ B(\VuHash,\leq)=B(\Vu,\geq)$.

\begin{example}
    For the CO system in Example~\ref{ex:main}, we have $\Core_G=B(\Vu,\geq)$ where
    \begin{align}
        B(\Vu,\geq)=\{ &r_V\in\RealP^{|V|}\colon r(\emptyset)=0, r(\{1\})\ge 1,               \nonumber \\
                       &r(\{2\})\ge 0, r\{3\})\ge 0,                 \nonumber \\
                       &r(\{1,2\})\ge 3, r(\{1,3\})\ge 3             \nonumber \\
                       &r(\{2,3\})\ge 1, r(\{1,2,3\})=\alpha   \}.   \nonumber
    \end{align}
    By converting constraint $r(X) \geq \Vu(X)$ to $r(V \setminus X)\leq \VuHash(V \setminus X)$ for all $X \subseteq V$, we have
    \begin{align}
        B(\VuHash,\leq)=\{ &r_V\in\RealP^{|V|} \colon r(\emptyset)=0, r(\{1\})\le \alpha-1,   \nonumber \\
                       &r(\{2\})\le \alpha-3, r(\{3\})\le \alpha-3,   \nonumber \\
                       &r(\{1,2\})\le \alpha, r(\{1,3\})\le \alpha   \nonumber \\
                       &r(\{2,3\})\le \alpha-1, r(\{1,2,3\})=\alpha   \}.  \nonumber
    \end{align}
    It can be shown that $ B(\VuHash,\leq)=B(\Vu,\geq),\forall\alpha\in\RealP$.

\end{example}

\subsection{Interpretation of Core in CO}

The core as expressed in \eqref{eq:Core} is in fact the set that contains all achievable rate vectors having sum-rate equal to $\alpha$ in CO. It can be interpreted as follows. Since $\Core_G=B(\VuHash,\leq)$, we can write $\Core_G$ as
\begin{multline}
    \Core_G = \Set{r_V\in\RealP^{|V|} \colon r(X) \leq \Vu(V)-\Vu(V \setminus X),\forall X \subseteq V,  \\
                    r(V)=\alpha}. \nonumber
\end{multline}
If $\Vu(X)$ is the payoff for forming coalition $X$, then $\Vu(V)-\Vu(V \setminus X)$ is the cost when coalition $X$ choose not to cooperate in the grand coalition $V$ \cite{Bilbao2012}.
A rate vector $r_V\in\RealP^{|V|}$ is a rate allocation method that divides sum-rate $r(V)=\alpha$ in $V$, where $r(X)$, the sum-rate in coalition $X$, can be considered as the cost for $X$ to be cooperative in $V$. Then, $r(X) \leq \Vu(V)-\Vu(V \setminus X)$ means that the cost when $X$ agrees to cooperate in $V$ is no greater than when $X$ denies to do so. If the core is nonempty, there exists at least one rate allocation method such that all users would not prefer forming the coalitions smaller than the grand one, or, alternatively speaking, there exists a rate allocation method that motivates all users to participate in information exchanging for CO. In this case, the game is called \textit{stable} \cite{Shapley1969Core} and the core can be considered as the solution set for the game.

\section{Nonemptiness of Core}

Since the core is not guaranteed to be nonempty in all coalitional games, there is a fundamental question:
\begin{enumerate}%[(a)]
    \item [(a)] When is the core nonempty?
\end{enumerate}
If the core is nonempty, we need to answer the question:
\begin{enumerate}
    \item [(b)] Can we find a rate vector $r_V$ in the core that allocates the value of the grand coalition $\Vu(V)=\alpha$ fairly among the users?
\end{enumerate}

The main purpose of this section is to answer question (a). We study the submodularity of the base polyhedron of the characteristic function to derive a necessary and sufficient condition for the nonemptiness of the core. Question (b) will be answered in Section~\ref{sec:Shapley}.

%\begin{definition}[sub/supermodular]
%    Function $g \colon 2^V \mapsto \Real$ is submodular if
%    \begin{equation}  \label{eq:SubMDef}
%        g(X) + g(Y) \geq g(X \cup Y) + g(X \cap Y), \quad \forall X, Y \subseteq V;
%    \end{equation}
%    $g$ is supermodular if $-g$ is submodular.
%\end{definition}
%
%In the following context, we show that all main results on coalitional game $G(V,\alpha,f_\alpha)$, e.g., the stability and convexity of the game and the nonemptiness of the core, are derived based on $\VuHash$.

\subsection{Necessary and Sufficient Condition}

Recall that $\Core_G=B(\Vu,\geq)=B(\VuHash,\leq)$, i.e., we can study either $B(\Vu,\geq)$ or $B(\VuHash,\leq)$ in order to determine the nonemptiness of $\Core_G$. In this section, we choose to consider $B(\VuHash,\leq)$ based on which we show that the condition for the nonemptiness of the core can be straightforwardly derived.

\begin{definition}[sub/supermodular]
    Function $g \colon 2^V \mapsto \Real$ is submodular if
    \begin{equation}  \label{eq:SubMDef}
        g(X) + g(Y) \geq g(X \cup Y) + g(X \cap Y), \quad \forall X, Y \subseteq V;
    \end{equation}
    $g$ is supermodular if $-g$ is submodular.
\end{definition}

\begin{lemma}
    If $\alpha \geq H(\RZ{V})$, $\VuHash$ is submodular, that is
    \begin{equation} \label{eq:SuMCond}
        \VuHash(X) + \VuHash(Y) \geq \VuHash(X \cup Y) + \VuHash(X \cap Y)
    \end{equation}
    for all $X,Y\subseteq V$; If $\alpha < H(\RZ{V})$, $\VuHash$ is intersecting submodular, that is inequality~\eqref{eq:SuMCond} holds for all $X,Y\subseteq V$ such that $X \cap Y \neq \emptyset$.
\end{lemma}
\begin{IEEEproof}
    For function $\VuHash$, we have
    \begin{align}
        &\quad \VuHash(X) + \VuHash(Y) - \VuHash(X \cup Y) - \VuHash(X \cap Y)=  \nonumber \\
        & \begin{cases}
                H(X) + H(Y) \\
                \quad - H(X \cup Y) - H(X \cap Y) + \alpha - H(V) & X \cap Y =\emptyset \\
                H(X) + H(Y) - H(X \cup Y) - H(X \cap Y)                 & \text{otherwise}
            \end{cases}.   \nonumber
    \end{align}
    Due to the submodularity of the entropy function $H$, i.e.,
    $$ H(X) + H(Y) \geq H(X \cup Y) + H(X \cap Y) \quad \forall X,Y \subseteq V,$$
    if $\alpha \geq H(\RZ{V})$, inequality~\eqref{eq:SuMCond} holds $\forall X,Y\subseteq V$, i.e., $\VuHash$ is submodular; if $\alpha < H(\RZ{V})$, inequality~\eqref{eq:SuMCond} holds $\forall X,Y\subseteq V \colon X \cap Y \neq \emptyset$, i.e., $\VuHash$ is intersecting submodular.
\end{IEEEproof}

Denote $\Pi(V)$ the partition set that contains all possible partitions of $V$ and $\Pi'(V)=\Pi(V)\setminus\{V\}$.

\begin{theorem} \label{theo:CoreNonEmpt}
    The core of $G(V,\alpha,\Vu)$ is nonempty if and only if
        \begin{equation} \label{eq:CoreNonEmptCond}
            \alpha = \min_{\Pat \in \Pi(V)} \sum_{C \in \Pat} \VuHash(C).
        \end{equation}
\end{theorem}
\begin{IEEEproof}
    Recall that $\Core_G=B(\VuHash,\leq)$. If $\alpha \geq H(\RZ{V})$, $\VuHash$ is submodular and $\VuHash(V)=\min_{\Pat \in \Pi(V)} \sum_{C \in \Pat} \VuHash(C)=\alpha$. Then, $B(\VuHash,\leq)$ is a submodular base polyhedron which is not empty \cite{Fujishige2005}. If $\alpha < H(\RZ{V})$, $\VuHash$ is intersecting submodular, and \eqref{eq:CoreNonEmptCond} is the necessary and sufficient condition for $B(\VuHash,\leq)$ to be nonempty according to Lemma~\ref{lemma:Main} in Appendix~\ref{app:lemma:Main}. Therefore, theorem holds.
\end{IEEEproof}

\subsection{Minimum sum-rate in CO}

Based on Theorem~\eqref{theo:CoreNonEmpt}, we can derive the minimum sum-rate for CO as follows.
\begin{corollary} \label{coro:MinSumRate}
    The core of $G(V,\alpha,\Vu)$ is non-empty if $\alpha\geq\RCO(\RZ{V})$, where
        \begin{equation} \label{eq:MinSumRateAsym}
            \RCO(\RZ{V}) = \max_{\Pat \in \Pi'(V)} \sum_{C \in \Pat} \frac{H(\RZ{V \setminus C}|\RZ{C})}{|\Pat|-1}
        \end{equation}
    for asymptotic model and
        \begin{equation} \label{eq:MinSumRateNonAsym}
            \RCO(\RZ{V}) = \Big\lceil \max_{\Pat \in \Pi'(V)} \sum_{C \in \Pat} \frac{H(\RZ{V \setminus C}|\RZ{C})}{|\Pat|-1} \Big\rceil
        \end{equation}
    for non-asymptotic model.
\end{corollary}
\begin{IEEEproof}
    Since $\VuHash(V)=\alpha$, \eqref{eq:CoreNonEmptCond} in Theorem~\ref{theo:CoreNonEmpt} is equivalent to $\alpha\leq\min_{\Pat \in \Pi'(V)} \sum_{C \in \Pat} \VuHash(C)$ which can be written as
        \begin{equation} \label{eq:IneqCoro}
            \alpha \geq \max_{\Pat \in \Pi'(V)} \sum_{C \in \Pat} \frac{H(\RZ{V \setminus C}|\RZ{C})}{|\Pat|-1}.
        \end{equation}
    Then, the minimum sum-rate must be the minimum value of $\alpha$ such that \eqref{eq:IneqCoro} holds. So, we have \eqref{eq:MinSumRateAsym} for asymptotic model. For non-asymptotic setting, $\alpha$ is the least integer number such that \eqref{eq:IneqCoro} holds. So, we have \eqref{eq:MinSumRateNonAsym}. According to Theorem~\ref{theo:CoreNonEmpt}, $\Core_G\neq\emptyset$ if $\alpha\geq\RCO(\RZ{V})$.
\end{IEEEproof}

\begin{figure}[tbp]
	\centering
    \scalebox{0.7}{% This file was created by matlab2tikz v0.4.3.
% Copyright (c) 2008--2013, Nico Schlömer <nico.schloemer@gmail.com>
% All rights reserved.
%
% The latest updates can be retrieved from
%   http://www.mathworks.com/matlabcentral/fileexchange/22022-matlab2tikz
% where you can also make suggestions and rate matlab2tikz.
%
%
% defining custom colors
\definecolor{mycolor1}{rgb}{1,1,0.5}%
\begin{tikzpicture}

\begin{axis}[%
width=3in,
height=2.5in,
area legend,
view={-33}{30},
scale only axis,
xmin=0,
xmax=4,
xlabel={\Large $r_1$},
xmajorgrids,
ymin=0,
ymax=1.5,
ylabel={\Large $r_2$},
ymajorgrids,
zmin=0,
zmax=1.5,
zlabel={\Large $r_3$},
zmajorgrids,
axis x line*=bottom,
axis y line*=left,
axis z line*=left,
legend style={at={(0.85,0.95)},anchor=north west,draw=black,fill=white,legend cell align=left}
]

\addplot3[solid,fill=mycolor1,draw=black]
table[row sep=crcr]{
x y z\\
3.2 0 0 \\
1.7 1.5 0 \\
0.2 1.5 1.5 \\
1.7 0 1.5 \\
3.2 0 0 \\
};
\addlegendentry{\Large $r(V)=3.2$};

\addplot3[solid,fill=white!90!black,opacity=4.000000e-01,draw=black]
table[row sep=crcr]{
x y z\\
0 0 0 \\
2.2 0 0 \\
2.2 0.2 0 \\
0 0.2 0 \\
0 0 0 \\
};
\addlegendentry{\Large $P(\VHash{3.2},\leq)$};

\addplot3[solid,fill=white!90!black,opacity=4.000000e-01,draw=black,forget plot]
table[row sep=crcr]{
x y z\\
0 0 0 \\
0 0.2 0 \\
0 0.2 0.2 \\
0 0 0.2 \\
0 0 0 \\
};

\addplot3[solid,fill=white!90!black,opacity=4.000000e-01,draw=black,forget plot]
table[row sep=crcr]{
x y z\\
0 0 0 \\
2.2 0 0 \\
2.2 0 0.2 \\
0 0 0.2 \\
0 0 0 \\
};

\addplot3[solid,fill=white!90!black,opacity=4.000000e-01,draw=black,forget plot]
table[row sep=crcr]{
x y z\\
2.2 0 0 \\
2.2 0.2 0 \\
2.2 0.2 0.2 \\
2.2 0 0.2 \\
2.2 0 0 \\
};

\addplot3[solid,fill=white!90!black,opacity=4.000000e-01,draw=black,forget plot]
table[row sep=crcr]{
x y z\\
0 0.2 0 \\
0 0.2 0.2 \\
2.2 0.2 0.2 \\
2.2 0.2 0 \\
0 0.2 0 \\
};

\end{axis}
\end{tikzpicture}%
}
	\caption{The the polyhedron $P(\VHash{3.2},\leq)$ for the CO system in Example~\ref{ex:main} and the plane $r(V)=3.2$. In this case, the intersection $\Core_G=B(\VHash{3.2},\leq)=\emptyset$. }
	\label{fig:DemoVuHash32}
\end{figure}

\begin{figure}[tbp]
	\centering
    \scalebox{0.7}{% This file was created by matlab2tikz v0.4.3.
% Copyright (c) 2008--2013, Nico Schlömer <nico.schloemer@gmail.com>
% All rights reserved.
%
% The latest updates can be retrieved from
%   http://www.mathworks.com/matlabcentral/fileexchange/22022-matlab2tikz
% where you can also make suggestions and rate matlab2tikz.
%
%
% defining custom colors
\definecolor{mycolor1}{rgb}{1,1,0.5}%
\definecolor{mycolor2}{rgb}{0.5,0.5,0.9}%
\begin{tikzpicture}

\begin{axis}[%
width=3in,
height=2.5in,
view={-33}{30},
scale only axis,
xmin=0,
xmax=4,
xlabel={\Large $r_1$},
xmajorgrids,
ymin=0,
ymax=1.5,
ylabel={\Large $r_2$},
ymajorgrids,
zmin=0,
zmax=1.5,
zlabel={\Large $r_3$},
zmajorgrids,
axis x line*=bottom,
axis y line*=left,
axis z line*=left,
legend style={at={(0.85,0.95)},anchor=north west,draw=black,fill=white,legend cell align=left}
]

\addplot3[area legend,solid,fill=mycolor1,draw=black]
table[row sep=crcr]{
x y z\\
3.5 0 0 \\
2 1.5 0 \\
0.5 1.5 1.5 \\
2 0 1.5 \\
3.5 0 0 \\
};
\addlegendentry{\Large $r(V)=3.5$};

\addplot3 [
color=mycolor2,
line width=6.0pt,
only marks,
mark=asterisk,
mark options={solid}]
table[row sep=crcr] {
2.5 0.5 0.5\\
};
\addlegendentry{\Large $B(\VHash{3.5},\leq)$};

\addplot3[area legend,solid,fill=white!90!black,opacity=4.000000e-01,draw=black]
table[row sep=crcr]{
x y z\\
0 0 0 \\
2.5 0 0 \\
2.5 0.5 0 \\
0 0.5 0 \\
0 0 0 \\
};

\addlegendentry{\Large $P(\VHash{3.5},\leq)$};

\addplot3[solid,fill=white!90!black,opacity=4.000000e-01,draw=black,forget plot]
table[row sep=crcr]{
x y z\\
0 0 0 \\
0 0.5 0 \\
0 0.5 0.5 \\
0 0 0.5 \\
0 0 0 \\
};

\addplot3[solid,fill=white!90!black,opacity=4.000000e-01,draw=black,forget plot]
table[row sep=crcr]{
x y z\\
0 0 0 \\
2.5 0 0 \\
2.5 0 0.5 \\
0 0 0.5 \\
0 0 0 \\
};

\addplot3[solid,fill=white!90!black,opacity=4.000000e-01,draw=black,forget plot]
table[row sep=crcr]{
x y z\\
2.5 0 0 \\
2.5 0.5 0 \\
2.5 0.5 0.5 \\
2.5 0 0.5 \\
2.5 0 0 \\
};

\addplot3[solid,fill=white!90!black,opacity=4.000000e-01,draw=black,forget plot]
table[row sep=crcr]{
x y z\\
0 0.5 0 \\
0 0.5 0.5 \\
2.5 0.5 0.5 \\
2.5 0.5 0 \\
0 0.5 0 \\
};

\end{axis}
\end{tikzpicture}%
}
	\caption{$B(\VHash{3.5},\leq)=\{(2.5,0.5,0.5)\}$ for the CO system in Example~\ref{ex:main}. In this system, $\RCO(\RZ{V})=3.5$ for asymptotic model. Consider game $G(V,3.2,\V{3.2})$. We have $\RRCO^*(\RZ{V})=\Core_G=B(\VHash{3.5},\leq)$ and $\{(2.5,0.5,0.5)\}$ is the only one optimal rate vector for CO.  }
	\label{fig:DemoVuHash35}
\end{figure}

\begin{example}
    For the CO model in Example~\ref{ex:main}, it can be show that $\RCO(\RZ{V})=3.5$ and $\RCO(\RZ{V})=4$ for asymptotic and non-asymptotic models, respectively, by applying \eqref{eq:MinSumRateAsym} and \eqref{eq:MinSumRateNonAsym}. If we increase $\alpha$ from $\alpha=0$ and observe the base polyhedron $B(\VuHash,\leq)$, the intersection of $P(\VuHash,\leq)$ and plane $r(V)=\alpha$, it can be shown that $P(\VuHash,\leq)$ does not intersect with plane $r(V)=\alpha$, i.e., $B(\VuHash,\leq)=\emptyset$, until $\alpha=3.5$. For example, $B(\VHash{3.2},\leq)=\emptyset$ in \fgref{fig:DemoVuHash32}. At $\alpha=3.5$, $B(\VHash{3.5},\leq)=\{(2.5,0.5,0.5)\}$ in \fgref{fig:DemoVuHash35} and $B(\VuHash,\leq)\neq\emptyset$ for all $\alpha\geq3.5$. But, $B(\VuHash,\leq)\cap\Z^3=\emptyset$ until $\alpha=4$ where $B(\VHash{4},\leq)\cap\Z^3=\{(3,0,1),(2,1,1),(3,1,0)\}$ as shown in Fig.~\ref{fig:DemoVuHash4}.
\end{example}

If we replace $H(\RZ{V \setminus C}|\RZ{C})$ with the cardinality function $|\Has{V}|-|\Has{C}|$ in \eqref{eq:MinSumRateAsym} and \eqref{eq:MinSumRateNonAsym}, we get exactly the minimum sum-rate expressions for asymptotic and non-asymptotic models, respectively, for CCDE in \cite{Ding2015ISIT}. Let $I(\RZ{V})$ be the MMI measure proposed in \cite{ChanMMI} that is defined as
$$ I(\RZ{V})=\min_{\Pat \in \Pi'(V)} \frac{D( P_{\RZ{V}} \| \prod_{C \in \Pat} P_{\RZ{C}} )}{|\Pat|-1}. $$
$D(\cdot \| \cdot)$ is the Kullback-Leibler divergence and $D( P_{\RZ{V}} \| \prod_{C \in \Pat} P_{\RZ{C}} )=\sum_{C\in\Pat}H(\RZ{C})-H(\RZ{V})$. We can write \eqref{eq:MinSumRateAsym} and \eqref{eq:MinSumRateNonAsym} as
\begin{equation} \label{eq:MinSumRateMMI}
    \RCO(\RZ{V})=H(\RZ{V})-I(\RZ{V})
\end{equation}
and
$$ \RCO(\RZ{V})=H(\RZ{V})- \lfloor I(\RZ{V}) \rfloor, $$
which are exactly the minimum sum-rate for CO for asymptotic and non-asymptotic models, respectively, in \cite{ChanSuccessive,ChanSuccessiveIT}. The interpretation of \eqref{eq:MinSumRateMMI} is as follows. $I(\RZ{V})$ can be considered as the maximum amount of information that is mutual to users in $V$ \cite{ChanMMI}. So, the minimum sum-rate for $\RCO(\RZ{V})$ must be $H(\RZ{V})-I(\RZ{V})$, the mount of information that is not mutual to users in $V$.

\begin{figure}[tbp]
	\centering
    \scalebox{0.7}{% This file was created by matlab2tikz v0.4.3.
% Copyright (c) 2008--2013, Nico SchlÃ¶mer <nico.schloemer@gmail.com>
% All rights reserved.
%
% The latest updates can be retrieved from
%   http://www.mathworks.com/matlabcentral/fileexchange/22022-matlab2tikz
% where you can also make suggestions and rate matlab2tikz.
%
%
% defining custom colors
\definecolor{mycolor1}{rgb}{0.5,0.5,0.9}%
\definecolor{mycolor2}{rgb}{1,1,0.5}%
\begin{tikzpicture}

\begin{axis}[%
width=3in,
height=2.5in,
view={-33}{30},
scale only axis,
xmin=0,
xmax=4,
xlabel={\Large $r_1$},
xmajorgrids,
ymin=0,
ymax=1.5,
ylabel={\Large $r_2$},
ymajorgrids,
zmin=0,
zmax=1.5,
zlabel={\Large $r_3$},
zmajorgrids,
axis x line*=bottom,
axis y line*=left,
axis z line*=left,
legend style={at={(0.95,0.95)},anchor=north west,draw=black,fill=white,legend cell align=left}
]

\addplot3[area legend,solid,fill=mycolor2,draw=black]
table[row sep=crcr]{
x y z\\
4 0 0 \\
2.5 1.5 0 \\
1 1.5 1.5 \\
2.5 0 1.5 \\
4 0 0 \\
};
\addlegendentry{\Large $r(V)=4$};

\addplot3[area legend,solid,fill=mycolor1,draw=black]
table[row sep=crcr]{
x y z\\
2 1 1 \\
3 0 1 \\
3 1 0 \\
2 1 1 \\
};
\addlegendentry{\Large $B(\VHash{4},\leq)$};

\addplot3[area legend,solid,fill=white!90!black,opacity=4.000000e-01,draw=black]
table[row sep=crcr]{
x y z\\
0 0 0 \\
3 0 0 \\
3 1 0 \\
0 1 0 \\
0 0 0 \\
};
\addlegendentry{\Large $P(\VHash{4},\leq)$};

\addplot3[solid,fill=white!90!black,opacity=4.000000e-01,draw=black,forget plot]
table[row sep=crcr]{
x y z\\
0 0 0 \\
0 1 0 \\
0 1 1 \\
0 0 1 \\
0 0 0 \\
};

\addplot3[solid,fill=white!90!black,opacity=4.000000e-01,draw=black,forget plot]
table[row sep=crcr]{
x y z\\
0 0 0 \\
3 0 0 \\
3 0 1 \\
0 0 1 \\
0 0 0 \\
};

\addplot3[solid,fill=white!90!black,opacity=4.000000e-01,draw=black,forget plot]
table[row sep=crcr]{
x y z\\
3 0 0 \\
3 1 0 \\
3 0 1 \\
3 0 0 \\
};

\addplot3[solid,fill=white!90!black,opacity=4.000000e-01,draw=black,forget plot]
table[row sep=crcr]{
x y z\\
0 1 0 \\
0 1 1 \\
2 1 1 \\
3 1 0 \\
0 1 0 \\
};

\addplot3[solid,fill=white!90!black,opacity=5.000000e-01,draw=black,forget plot]
table[row sep=crcr]{
x y z\\
0 0 1 \\
3 0 1 \\
2 1 1 \\
0 1 1 \\
0 0 1 \\
};

\addplot3 [
color=red,
line width=3.0pt,
only marks,
mark=triangle,
mark options={solid,,rotate=180}]
table[row sep=crcr] {
2 1 1\\
3 0 1\\
3 1 0\\
2 1 1\\
};
\addlegendentry{\Large $B(\VHash{4},\leq)\cap\Z^3$};

\addplot3 [
color=black,
line width=5.0pt,
only marks,
mark=star,
mark options={solid}]
table[row sep=crcr] {
2.66666666666667 0.666666666666667 0.666666666666667\\
};
\addlegendentry{\Large Shapley value};

\end{axis}
\end{tikzpicture}%
}
	\caption{$B(\VHash{4},\leq)$ for the CO system in Example~\ref{ex:main}. In this system, $\RCO(\RZ{V})=4$ for non-asymptotic model. Consider game $G(V,4,\V{4})$. We have $\RRCO^*(\RZ{V})\cap\Z^3=\Core_G\cap\Z^3=B(\VHash{4},\leq)\cap\Z^3=\{(3,0,1),(2,1,1),(3,1,0)\}$. The Shapley value calculated in Example~\ref{ex:Shapley} lies in the core.}
	\label{fig:DemoVuHash4}
\end{figure}

\subsection{Convexity of Game} \label{sec:ConvexGame}

Convex game is a special class of coalitional game.
\begin{definition}[Convex Game \cite{Shoham2008}]
    A coalitional game is convex if the characteristic function is supermodular.
\end{definition}
Convex game has nice properties \cite{Sapley1971Convex}:
\begin{itemize}
    \item The core is nonempty;
    \item Shapley value lies in the core.
\end{itemize}
In this section, we use the first property to interpret Corollary~\ref{coro:MinSumRate}. The second property will be used to present a fair distribution of the value of the grand coalition $f_\alpha(V)=\alpha$ that lies in the core in Section~\ref{sec:Shapley}.

\begin{lemma} \label{lemma:Convex}
    For each $\alpha\geq\RCO(\RZ{V})$, there exists a convex game $\hat{G}(V,\alpha,\VuHat)$ such that the cores of $G$ and $\hat{G}$ are equal.
\end{lemma}
\begin{IEEEproof}
    The Dilworth truncation of $\VuHash$ is given by \cite{Dilworth1944}
    \begin{equation} \label{eq:Dilworth}
        \VuHashHat(X)=\min_{\Pat\in \Pi(X)} \sum_{C\in\Pat} \VuHash(C), \forall X \subseteq V.
    \end{equation}
    For $\alpha\geq H(\RZ{V})$, $\VuHashHat=\VuHash$ since $\VuHash$ is submodular; For $\RCO(\RZ{V})\leq\alpha< H(\RZ{V})$, $\VuHashHat$ is submodular, $\VuHashHat(V)=\VuHash(V)=\alpha$ and $B(\VuHashHat,\leq)=B(\VuHash,\leq)$ \cite{Fujishige2005}. Let
    $$ \VuHat(X) = \VuHashHat(V) - \VuHashHat(V \setminus X). $$
    $\VuHat$ is supermodular and $B(\VuHat,\geq)=B(\VuHash,\leq)$. Here, $B(\VuHat,\geq)$ is the core of game $\hat{G}(V,\VuHat,\alpha)$. Recall that $\Core_G=B(\VuHash,\leq)$. Therefore, $\Core_G=\Core_{\hat{G}}$. In addition, if $\VuHash$ is integer-valued, so is $\VuHashHat$ \cite{Fujishige2005}. Therefore, $\Core_G\cap\Z^{|V|}=\Core_{\hat{G}}\cap\Z^{|V|}$ for non-asymptotic model.
\end{IEEEproof}

We use Lemma~\ref{lemma:Convex} to interpret Corollary~\ref{coro:MinSumRate} as follows. If $\alpha\geq\RCO(\RZ{V})$, game $G(V,\alpha,\Vu)$ is equivalent to convex game $\hat{G}(V,\alpha,\VuHat)$, where $\VuHat$ can be obtained from $\Vu$ by Dilworth truncation. In this case, $\Core_G=\Core_{\hat{G}}\neq\emptyset$ for asymptotic model and $\Core_G\cap\Z^{|V|}=\Core_{\hat{G}}\cap\Z^{|V|}\neq\emptyset$ for non-asymptotic model due to the convexity of $\hat{G}$.

Since function $\VuHashHat$ is submodular, the optimization problem on the core $\Core_G=B(\VuHashHat,\leq)$ is closely related to SFM that can be solved in strongly polynomial time. For example, the problem of finding a rate vector $r_V\in\Core_G=B(\VuHashHat,\leq)$ can be solved by Edmond's greedy algorithm \cite{Edmonds2003Convex}. Related algorithms can be found in \cite{CourtIT2014,MiloIT2015}. These algorithms always find an extreme point (a vertex) of the core.

\begin{example} \label{ex:Convex}
    Consider the CO system in Example~\ref{ex:main} when $\alpha=4$ and its corresponding game model $G(V,4,\V{4})$. We obtain $\VHashHat{4}$ by the Dilworth truncation~\eqref{eq:Dilworth} and obtain $\VHat{4}$ as
    \begin{align}
        &\VHat{4}(\emptyset)=0,\VHat{4}(\{1\})=2,\VHat{4}(\{2\})=0,\VHat{4}(\{3\})=0, \nonumber \\
        &\VHat{4}(\{1,2\})=3,\VHat{4}(\{1,3\})=3,\VHat{4}(\{2,3\})=1,   \nonumber \\
        &\VHat{4}(\{1,2,3\})=4.
    \end{align}
    $\VHat{4}$ is supermodular and therefore $\hat{G}(V,4,\VHat{4})$ is a convex game. It can be shown that $\Core_G=B(\V{4},\geq)=B(\VHat{4},\geq)=\Core_{\hat{G}}$ and $\Core_G\cap\Z^{|V|}=B(\V{4},\geq)\cap\Z^{|V|}=B(\VHat{4},\geq)\cap\Z^{|V|}=\Core_{\hat{G}}\cap\Z^{|V|}$. In addition, $B(\VHat{4},\geq)=B(\VHashHat{4},\leq)=B(\VHash{4},\leq)$, where $B(\VHash{4},\leq)$ is shown in \fgref{fig:DemoVuHash4}.
\end{example}

\section{Shapley Value}  \label{sec:Shapley}

To make sure that all users are willing to cooperate, it is not sufficient to just know that the core is nonempty. Since the core is not a singleton in general, some solutions in the core may not be a good choice in terms of fairness. For example, for the CO system in Example~\ref{ex:main} when $\alpha=4$, the Edmond's greedy algorithm usually returns a vertex of the core $\Core_G=B(\VHash{4},\leq)$, e.g., $(3,0,1)$, which is one of the unfairest solutions in $\Core_G$ that may make one or more users unwilling to cooperate. In CCDE, to achieve some degree of fairness in rate allocation can also prevent running out of mobile clients' battery usage.

The authors in \cite{MiloFair2012,Ding2015ICT} proposed polynomial time algorithms to compute a fair rate vector in core $\Core_G$ for non-asymptotic model in CCDE. The greedy algorithm in \cite{MiloFair2012} is based on SFM. The authors in \cite{Ding2015ICT} have shown that a fair rate in the base polyhedron $B(\VuHash,\leq)$, or the core $\Core_G$, can be found by solving an $M$-convex minimization for which there exists a discrete steepest descent algorithm that can search the optimal solution in polynomial time.

Finding a fair rate vector for the asymptotic model is more complex than non-asymptotic one. Although the problem can be easily formulated by a convex minimization problem, e.g., $\min\{\sum_{i\in V} r_i^2 \colon r_V\in\Core_G\}$, where the objective function is defined based on Jain's fairness index \cite{Jain1984}, the number of constrains $2^{|V|}$ is exponentially growing in $|V|$. In \cite{Taj2011}, the authors build a multi-layer graph model and formulate a convex minimization problem based on it. Although the constraints is not as large as $2^{|V|}$, building new model and deriving constraints based on this model incurs extra complexity. In this section, we show an alternative way to achieve the fairness based on game model $G(V,\alpha,\Vu)$: Shapley value.

\begin{definition}[Shapley Value \cite{Saad2009Game}]
    For a coalitional game with characteristic function $g \colon 2^V \mapsto \Real$, the Shapley Value is a rate vector $\hat{r}_V$ with each entry being
        $$ \hat{r}_i = \sum_{X \subseteq V \colon i \in X} \frac{ (|X|-1)! |V\setminus{X}|! }{|V|!} \Big( g(X)-g(X\setminus{\{i\}}) \Big).$$
\end{definition}

The weight factor $\frac{ (|X|-1)! |V\setminus{X}|! }{|V|!}$ is the probability for user $i$ to enter coalition $X$ in a random order \cite{Saad2009Game}. Therefore, $\hat{r}_i$ can be considered as the expected marginal value of the characteristic function $g$ when the users join coalitions randomly to form the grand coalition $V$ \cite{Shapley1969Core}.

\begin{lemma} \label{lemma:Shapley}
    For asymptotic model, if $\alpha\geq\RCO(\RZ{V})$, the Shapley value $\hat{r}_V$ with
        $$ \hat{r}_i = \sum_{X \subseteq V \setminus \{i\}} \frac{ (|V\setminus{X}|-1)! |X|! }{|V|!} \Big( \VuHashHat(X \cup \{i\})-\VuHashHat(X) \Big) $$
    lies in the core and is a fair method to distribute the value of the grand coalition $\Vu(V)=\alpha$ among all users.
\end{lemma}
\begin{IEEEproof}
    Based on the proof of Lemma~\ref{lemma:Convex}, if $\alpha\geq\RCO(\RZ{V})$, $\Core_G=\Core_{\hat{G}}$. Game $\hat{G}(V,\alpha,\VuHat)$ is convex and the characteristic function $\VuHat$ is defined as $\VuHat(X) = \VuHashHat(V) - \VuHashHat(V \setminus X),\forall X \subseteq V$. The Shapley value of $\hat{G}$ is
        \begin{align}
            \hat{r}_i &= \sum_{X \subseteq V \colon i \in X} \frac{ (|X|-1)! |V\setminus{X}|! }{|V|!} \Big( \VuHat(X)-\VuHat(X\setminus{\{i\}}) \Big)   \nonumber \\
                      &= \sum_{X\subseteq V \setminus \{i\}} \frac{ (|V \setminus X|-1)! |X|! }{|V|!} \Big( \VuHashHat(X\cup\{i\}) - \VuHashHat(X) \Big).   \nonumber
        \end{align}
    It is shown in \cite{Lan2010Fair} that $\hat{r}_V\in\Core_{\hat{G}}$ and is a fair method to distribute $\alpha$ in set $V$.
\end{IEEEproof}

\begin{example} \label{ex:Shapley}
    Consider the CO system in Example~\ref{ex:main}. We obtain function $\VHashHat{4}$ by Dilworth truncation~\eqref{eq:Dilworth} as
    \begin{align}
        &\VHashHat{4}(\emptyset)=0,\VHashHat{4}(\{1\})=3,\VHashHat{4}(\{2\})=1,\VHashHat{4}(\{3\})=1, \nonumber \\
        &\VHashHat{4}(\{1,2\})=4,\VHashHat{4}(\{1,3\})=4,\VHashHat{4}(\{2,3\})=2,   \nonumber \\
        &\VHashHat{4}(\{1,2,3\})=4.  \nonumber
    \end{align}
    Note, the values of $\VHashHat{4}$ have been used to determine function $\VHat{4}$ in Example~\ref{ex:Convex}. For user 1, the possible values of $X\subseteq V\setminus\Set{1}$ are $\emptyset$, $\{2\}$, $\{3\}$ and $\{2,3\}$. Therefore,
    \begin{align}
        \hat{r}_1 = & \frac{2!0!}{3!}(3-0) + \frac{1!1!}{3!}(4-1) \nonumber \\
                    & + \frac{1!1!}{3!}(4-1) + \frac{0!2!}{3!}(4-2) =\frac{8}{3}.
    \end{align}
    In the same way, we find $\hat{r}_2=\frac{2}{3}$ and $\hat{r}_3=\frac{2}{3}$. The Shapley value $\hat{r}_V=(\frac{8}{3},\frac{2}{3},\frac{2}{3})$ is plotted in \fgref{fig:DemoVuHash4}, where it can be seen that $\hat{r}_V\in\Core_G$ and $\hat{r}_V$ is a fair rate allocation method as compared to other rate vectors in the core.
    %$\hat{r}_V$ is also the center of gravity of $\Core_G$. Consider the three extreme points/vertex $(3,0,1)$, $(2,1,1)$ and $(3,1,0)$, $\hat{r}_V=\frac{(3,0,1)+(2,1,1)+ (3,1,0)}{3}=(\frac{8}{3},\frac{2}{3},\frac{2}{3})$.
\end{example}

Although obtaining the Dilworth truncation $\hat{f}^\#(X)$ for a particular $X$ is related to SFM and can be completed in strongly polynomial time \cite{Schrijver2003}, the complexity of obtaining Shapley value is exponentially growing with $|V|$. How to reduce the complexity of calculating Shapley value is still an open problem in the literature \cite{Saad2009Game}. However, the advantage of Shapley value over the method in \cite{Taj2011} is that it can be learned in a decentralized manner: As long as user $i$ obtains the knowledge of the whole system, e.g., $\Vu$, his/her tuple $\hat{r}_i$ in Shapley value can be calculated by him/herself.

%\Ni{Shapley value is a fair solution with respect to certain fairness axiom: symmetry, dummy and decomposable. I'm avoid using fairest since a 'fairest' solution based on axiom may be unfair on another one. There are may fairness axioms in the literature \cite{Taj2011}. I am not sure whether Jain's index fits to the axiom of Shapley value. I also do not have the proof that the fairest in Jain's index equal to Shapley value at hand, although it is the case in Example~\ref{ex:main}. In addition, I'm still at the entry level to understand coalitional game model. This paper is just a taste of game model.}
%

\section{Conclusion}
We formulated the problem of CO by a coalition game model where the core contained all achievable rate vectors that satisfied the Slepian-Wolf constraints for CO and had sum-rate equal to a given value. We showed that the core was a base polyhedron of a submodular or intersecting submodular function. We derived the necessary and sufficient condition for the core to be nonempty, based on which we gave the expression of the minimum sum-rate for CO and showed that they were consistent with the existing results in the literature. We proved that the game was convex when the sum-rate was greater or equal to the minimum sum-rate and showed that a fair rate vector in the core could be obtained by calculating the Shapley value. However, since the complexity of obtaining the Shapley value is exponentially growing, it is worth discussing how to allow users to learn the fair rate vector in the core in polynomial time, which could be one of the direction of research work in the future.

%\newpage

\appendices

\section{} \label{app:lemma:Main}
%{Lemma~\ref{lemma:Main}}
%Let $g \colon 2^V \mapsto \RealP$ be an intersecting submodular function. We call $B(g,\leq)$ intersecting submodular base polyhedron.
\begin{lemma} \label{lemma:Main}
    For an intersecting submodular function $g \colon 2^V \mapsto \RealP$ such that $g(\emptyset)=0$, the base polyhedron $B(g,\leq)$ is nonempty if and only if
    $$ g(V) = \min_{P \in \Pi(V)} \sum_{C \in P} g(C) $$
\end{lemma}
\begin{IEEEproof}
    Theorem 2.6 in \cite{Fujishige2005} gives the necessary and sufficient conditions for $B(g,\leq)$ to be nonempty:
            \begin{equation} \label{eq:FujishCond}
                g(V)=\max_{P\in\Pi(V)} \sum_{C \in P} g^\#(C)=\min_{P\in\Pi(V)} \sum_{C \in P} g(C),
            \end{equation}
    which is equivalent to $g(V) \geq \sum_{C \in P} g^\#(C)$ and $g(V) \leq \sum_{C \in P} g(C)$ for all $P \in \Pi'(V)$. The latter can be written as
            \begin{align}
                &\quad g(V) \leq \sum_{C \in P} ( g(V) - g^\#(V \setminus C) )  \nonumber \\
                &\Rightarrow g(V) \leq |P| g(V) - \sum_{C \in P} g^\#(V \setminus C)  \nonumber \\
                &\Rightarrow g(V) \geq \frac{\sum_{C \in P} g^\#(V \setminus C)}{|P|-1}, \quad \forall P \in \Pi'(V) .  \nonumber
            \end{align}
    Due to the intersecting submodularity of $g$, for all $\emptyset\neq X,Y \subset V$ such that $X \cap Y = \emptyset$ and $X \cup Y \neq V$, we have
    $$ g^\#( X \cup Y) \geq g^\#(X) + g^\#(Y). $$
    So, for all $C \in P$ where $P$ is some partition in $\Pi'(V)$, $ g^\#(V \setminus C) \geq \sum_{C' \in P \setminus C} g^\#(C')  $ and
    $ \sum_{C \in P}g^\#(V \setminus C) \geq (|P|-1) \sum_{C \in P } g^\#(C),$
    i.e.,
    $$ \sum_{C \in P} \frac{g^\#(V \setminus C)}{|P|-1} \geq \sum_{C \in P } g^\#(C), \quad \forall P \in \Pi'(V).  $$
    Therefore, \eqref{eq:FujishCond} reduces to $g(V) \leq \sum_{C \in P} g(C),\forall P \in \Pi'(V)$, which is equivalent to $g(V) = \min_{P \in \Pi(V)} \sum_{C \in P} g(C)$.
\end{IEEEproof}

\bibliographystyle{ieeetr}
\bibliography{CDE_CCDE_BIB}

\end{document}